\documentclass[superscriptaddress,prl,nobalancelastpage,tightenlines,nofootinbib]{revtex4-2}
\usepackage{amssymb}
\usepackage{amsmath}
\usepackage{array}
\usepackage{graphicx}
\usepackage{tikz}
\usepackage{pgfplots}
\usepackage{placeins}
\usepackage{relsize}
\usepackage{tabu,multirow}
\usepackage{epsfig}
\usepackage{latexsym}
\usepackage{subfigure}
\usepackage{soul}
\usepackage[export]{adjustbox}
\usepackage{setspace}
\usepackage{float}
\usepackage{stmaryrd}
\usepackage{miller}
\usepackage{bm}
\usepackage{tabularx}
\usepackage{booktabs}
\usepackage[hidelinks]{hyperref} 
\usepackage[noabbrev,capitalize]{cleveref}
\usepackage{ragged2e}
\usepackage{titlesec}
\usepackage{multirow}

\usepackage{pdfpages} 
\usepackage{pgffor} 

\makeatletter
\AtBeginDocument{\let\LS@rot\@undefined}
\makeatother

\def\supplementfilename{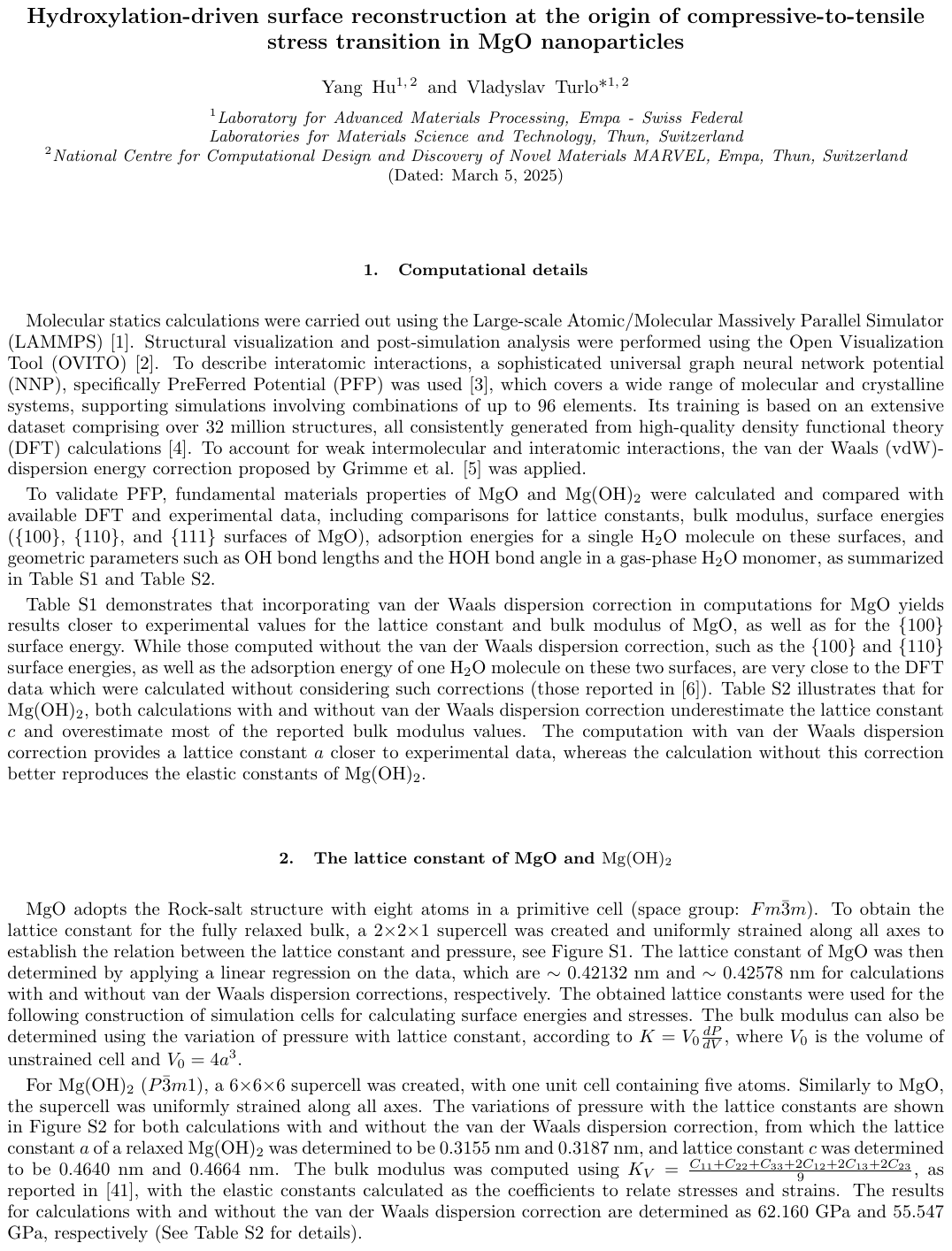}

\pdfximage{\supplementfilename}
\def\numbersupplementpages{\the\pdflastximagepages}

\newif\ifarXiv
\arXivtrue 

\begin{document}

\title{Hydroxylation-driven surface reconstruction at the origin of compressive-to-tensile stress transition in metal oxide nanoparticles}

\author{Yang Hu}
\affiliation{Laboratory for Advanced Materials Processing, Empa - Swiss Federal Laboratories for Materials Science and Technology, Thun, Switzerland\\}%
\affiliation{National Centre for Computational Design and Discovery of Novel Materials MARVEL, Empa, Thun, Switzerland\\}%

\author{Vladyslav Turlo*}
\affiliation{Laboratory for Advanced Materials Processing, Empa - Swiss Federal Laboratories for Materials Science and Technology, Thun, Switzerland\\}%
\affiliation{National Centre for Computational Design and Discovery of Novel Materials MARVEL, Empa, Thun, Switzerland\\}%
\email{vladyslav.turlo@empa.ch}
\date{\today}

\maketitle
\section*{Abstract}
Experiments reveal negative (non-Laplacian) surface stresses in metal oxide nanoparticles, partly associated with humidity during fabrication and annealing. Using a neural network interatomic potential for MgO, we prove that water adsorption induces surface hydroxylation, shifting facets from \{100\} to \{110\} to \{111\} and switching the average surface stress from positive to negative. Predicted lattice strains versus nanoparticle size agree well with experiments, clarifying experimental correlations. The new framework informs broad applications in catalysis, sensors, batteries, and biomedicine.

\section*{Main text}
Nanoparticles, typically defined as particles with dimensions on the order of 100 nm or less, exhibit properties significantly different from those of bulk materials. These differences arise primarily due to their high surface-to-volume ratio and the quantum mechanical effects associated with their nanoscale dimensions \cite{diehm2012size,santos2015industrial,sajid2020nanoparticles}. As a result, nanoparticles have unique mechanical, electrical, optical, magnetic, and chemical reactivity properties, making them highly versatile for applications in catalysis, quantum dots, sensors, batteries, biomedicine, and more \cite{stark2015industrial,santos2015industrial,fahmy2022review,hornak2021synthesis,sajid2020nanoparticles,du2022identification,shanawad2023humidity,asha2024improved,wang2019enhancing,pandey2009magnesium,kumar2007effect,stoimenov2002metal,huang2005controllable,applerot2009enhanced,cai2017highly,behzadi2019albumin}.

Key parameters used to characterize nanoparticles include their crystal structure (e.g., composition, lattice constant(s), space group), surface area, surface chemistry, surface charge, and morphology \cite{santos2015industrial,sajid2020nanoparticles}. These properties are critical for the physical and chemical behaviors of nanoparticles. One particularly intriguing phenomenon, observed since the 1960s, is the size-dependent variation in lattice strain in nanoparticles. For instance, noble metal nanoparticles made of platinum \cite{wasserman1972determination}, silver \cite{wasserman1970determination}, and gold \cite{mays1968surface} demonstrate a reduction in lattice size as particle size decreases.

This size-dependent lattice contraction has been attributed to positive surface stresses in metallic nanoparticles, as confirmed by \textit{ab initio} calculations \cite{diehm2012size}. 
A quantitative relation between the lattice strain $\frac{\Delta a}{a_0}$ and the surface stress $f$ can be derived from the Laplace pressure by assuming a spherical nanoparticle \cite{rodenbough2017lattice,chan2021surface}: 
\begin{equation}
    \frac{\Delta a}{a_0}=-\frac{4f}{3Bd}
\label{eq:stre_size}
\end{equation}
where $a_0$ is the bulk lattice constant, $\Delta a$ is the difference between the lattice constant measured for nanoparticles and the bulk lattice constant, $B$ is the bulk modulus, and $d$ is the particle size. Eq. (\ref{eq:stre_size}) shows the opposite sign between the surface stress and lattice strain, and an inverse relation between the lattice strain and particle size.

In contrast to noble metals, many metal oxide nanoparticles exhibit either lattice contraction or expansion as particle size decreases \cite{zhou2001size,chen2010size,prieur2020size,rodenbough2017lattice,guilliatt1969lattice,cimino1966dependence,li2004evidence,li2007p,ali2010zno,okubo2007nanosize,su2007tunable,bhowmik2006lattice,wu2007negative,naughton2007lattice}. For example, molecular dynamics simulations of $\gamma$-$\text{Al}_2\text{O}_3$ nanoparticles particles annealed in a vacuum at 900 K revealed lattice expansion in the core region, accompanied by surface amorphization and Al segregation to the surface, suggesting a strong correlation between changes in surface chemistry and lattice expansion \cite{gramatte2023atomistic}. Similarly, in CeO$_2$ nanoparticles, the increase in lattice constant was linked to the presence of excess oxygen vacancies at the surface and the associated reduction in Ce valence \cite{zhou2001size,chen2010size}, also confirmed with \textit{ab initio} calculations\cite{loschen2008density}. 

Surface chemistry, however, is not only modified by defect formation or cation segregation but also by environmental factors, such as water adsorption. Experiments on MgO nanoparticles demonstrated an increased lattice constant for particles exposed to moist air compared to those maintained under vacuum \cite{guilliatt1969lattice,cimino1966dependence}. These findings suggest that water adsorption contributes to lattice expansion, although this hypothesis remains unproven. Applying the established relation between lattice strain and surface stress (Eq. (\ref{eq:stre_size})), a negative surface stress would be required to induce such lattice expansion. While prior studies have explored the impact of surface hydroxylation on surface energies \cite{lodziana2004negative,geysermans2009combination}, its effect on surface stress has not been investigated in the literature.

To address this gap, molecular statics simulations were performed here to evaluate the surface stresses of MgO nanoparticles under varying levels of water coverage. MgO was selected as the model system due to its simple rock salt structure and its experimentally observed propensity for water adsorption \cite{tang2014analysis}. 
Our simulations revealed that among the three low-index MgO surfaces, the \{100\} surface exhibits positive surface stresses while the \{110\} and \{111\} surfaces show negative surface stresses regardless of water coverage, making hydroxylation-induced surface reconstruction the key reason for compressive-to-tensile stress transition in MgO nanoparticles. The predicted range of lattice strains with respect to nanoparticle sizes is in excellent agreement with the experimental data, highlighting the reliability of our computational analysis framework.

In this work, we use the Large-scale Atomic/Molecular Massively Parallel Simulator (LAMMPS) \cite{Plimpton1995} for molecular statics calculations and the Open Visualization Tool (OVITO)  \cite{Stukowski2010} for analysis and visualization. We chose a universal graph neural network potential (PreFerred Potential, PFP) \cite{takamoto2022towards} to describe interatomic interactions, due to the extensive fitting database of over 50 million carefully curated \textit{ab initio}-computed structures covering 96 elements of the periodic table. The potential has been successfully applied to study similar systems \cite{mine2023comparison}, for example, uncovering hydrogen distribution in amorphous alumina \cite{Gramatte2024}. In our tests, the PFP demonstrated strong agreement with available experimental and density functional theory (DFT) data for the fundamental properties of Mg-O-H systems (See Tables S1 and S2), especially while incorporating van der Waals corrections \cite{Grimme2010} to account for potential weak interactions between water molecules and the MgO surface. In molecular statics calculations, atomic positions were optimized using first the conjugate gradient (CG) energy minimization method \cite{berne1998classical}, followed by the Fast Inertial Relaxation Engine (FIRE) method \cite{bitzek2006structural,guenole2020assessment}, while keeping the simulation box dimensions fixed. This ensures the complete relaxation of atomic positions and accurate results for computed surface energies. 

The surface stress $f(\varepsilon)$ under a biaxial in-plane strain, $\varepsilon$, is related to the surface energy, $\gamma(\varepsilon)$, by the Shuttleworth equation:
\begin{equation} 
f(\varepsilon)=\gamma(\varepsilon)+\frac{\partial{\gamma(\varepsilon)}}{\partial{\varepsilon}},
\label{eq:interfstre}
\end{equation} 
thus requiring us to compute the dependence of surface energy on strain for all considered surfaces. As we have recently shown, the derivative of surface energy on strain is decisive for determining the sign and magnitude of surface/interface stress \cite{Lorenzin2024,Hu2024}. 
The surface energy can be computed as follows:
\begin{equation}    
\gamma(\varepsilon)=\frac{E_{tot}^{surf}(\varepsilon)-E_{tot}^{bulk}(\varepsilon)-n\times E_{\mathrm{H_2O}}}{2A(\varepsilon)}
\label{eq:interfen}
\end{equation}
where $E_{tot}^{surf}(\varepsilon)$ and $E_{tot}^{bulk}(\varepsilon)$ are the potential energies of simulation cells with and without surfaces at biaxial in-plane strain, $\varepsilon$, and $2A$ is the total surface area at the same strain. $n$ is the number of $\mathrm{H_2O}$ molecules and $E_{H_{2}O}$ is the energy of a $\mathrm{H_2O}$ monomer in vacuum. In such case, we consider water pressure to be negligible, so only molecules adsorbed to the surface are affecting its energy.

\begin{figure}[htbp!]
  \centering
  \includegraphics[width=0.95\textwidth,clip,trim=0cm 0.45cm 0cm 0cm]{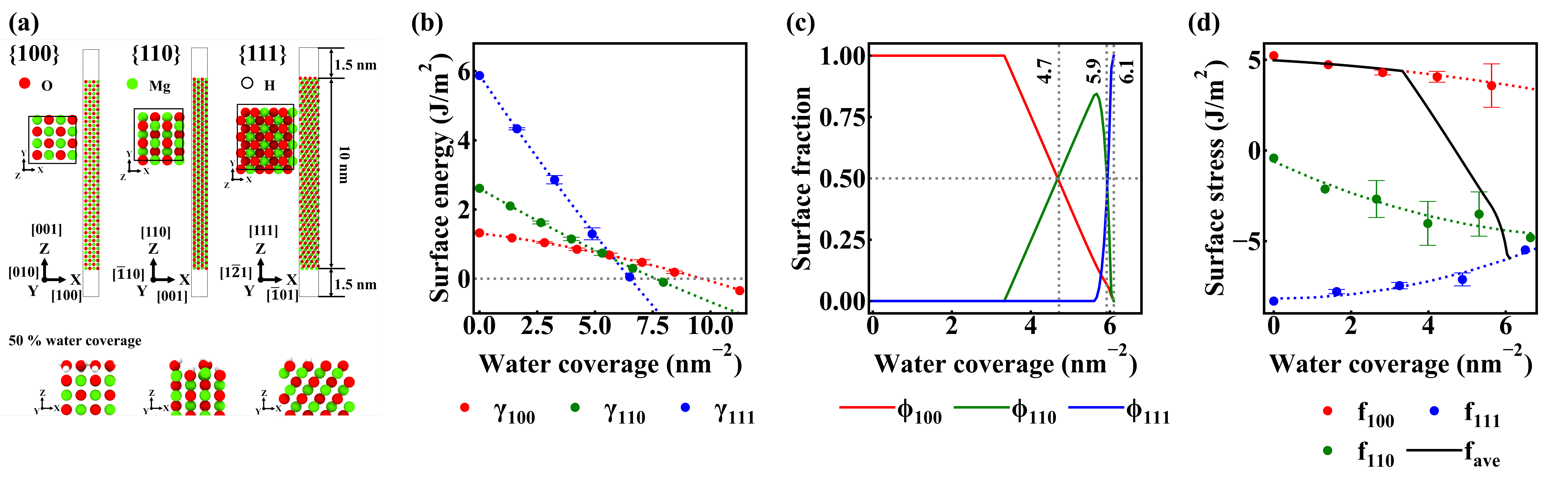}
  \caption{(a) MD simulation cells for the \{100\}, \{110\} and \{111\} surfaces of MgO, with the side views of water molecules on each surface of MgO for 50\% surface water coverage. Mg atoms are colored green, oxygen atoms are colored red, and hydrogen atoms are colored white. (b) The variation of surface energies with the number of $\mathrm{H_2O}$ molecules per surface area for the \{100\}, \{110\} and \{111\} surfaces of MgO. Dotted lines are obtained by fitting the data of each surface using quadratic regression. (c) The variation of surface fractions with the number of $\mathrm{H_2O}$ molecules per surface area for the \{100\}, \{110\} and \{111\} surfaces of MgO. Grey dotted lines show the water concentration on the surface at 50\% of \{110\}, and 50\% of \{111\} surfaces. (d) The variation of surface stresses with the number of $\mathrm{H_2O}$ molecules per surface area for the \{100\}, \{110\} and \{111\} surfaces of MgO. Dotted lines are obtained by fitting the data of each surface using quadratic regression. The black solid line is the average surface stress of a nano-particle, computed using the surface fraction and surface stress of each surface in (c) and (d).}
  \label{surf_frac}
\end{figure}

After testing and mitigating the size effects, slabs for the three low-index surfaces of MgO (\{100\}, \{110\}, \{111\}) were set up as shown in Fig. \ref{surf_frac}(a). Then, water molecules were introduced by adding pairs of OH groups and H ions to the top and bottom surfaces attached to neighboring Mg and O surface atoms, respectively. If dissociation on a particular surface is unfavorable, OH and H would be recombined to form water molecules after structural relaxation. Water coverages ranging from 0 to 100\% with various spatial arrangements of water molecules were tested (Figs. S6-S8). Physisorption of $\mathrm{H_2O}$ molecules was observed on the \{100\} surface. While on the \{110\} surface, water molecules always dissociated regardless of the level of water coverage, indicating consistent chemisorption, being consistent with \cite{chen2023hydration}. For the \{111\} surface, OH groups and H atoms were introduced separately to the Mg-terminated and O-terminated surfaces, respectively, which hugely reduces the surface energy. Although this separation was imposed artificially, such distribution of OH and H on Mg- and O-terminated surfaces could naturally occur on nanoparticles due to surface diffusion.

Computed surface energies as functions of water coverage (in terms of $\mathrm{H_2O}$ molecules per surface area) are shown in Fig. \ref{surf_frac}(b). All three surfaces exhibit a decreasing trend in surface energy with increasing water coverage, indicating that water adsorption on the surface is energetically favorable. For both \{100\} and \{110\} surfaces, negative surface energies are obtained for surface water coverage of 100\% (-0.34 J/m$^2$ and -0.10 J/m$^2$, respectively). While for the \{111\} surfaces, full water coverage results in a near zero surface energies (0.05 J/m$^2$). Similar observations of negative surface energies have been reported for hydroxylated surfaces in $\theta-\text{Al}_2\text{O}_3$ \cite{lodziana2004negative}, and was believed to explain the strong stability of such phases upon high temperatures. 

Changes in surface energies directly affect the morphology of nanoparticles, i.e., the relative surface areas of different facets. According to the theory of Wulff construction \cite{dobrushin1992wulff}, the ratio of surface energy to the distance from the particle center to the surface center in a fully equilibrated nanoparticle remains constant. Therefore, the variations in surface energy with water coverage, as shown in Fig. \ref{surf_frac}(a), were fitted using quadratic regressions for all three surfaces, and then used as inputs into the WulffPack code \cite{rahm2020wulffpack} to calculate the surface area fractions at specific water coverage levels, provided in Fig. \ref{surf_frac}(b).

Initially, nanoparticles are entirely enclosed by \{100\} surfaces, up to a water coverage of approximately 3.3 molecules/nm$^2$. Beyond this threshold, \{110\} surfaces begin to appear, with their fraction increasing as water coverage rises, reaching an equal fraction with \{100\} surfaces at approximately 4.7 molecules/nm$^2$. At around 5.6 molecules/nm$^2$, \{111\} surfaces start to emerge and quickly dominate due to their lower surface energies at high water coverage. A nanoparticle composed of 50\% \{111\} surface and 50\% of the other two surfaces can be achieved at approximately 5.9 molecules/nm$^2$. All curves terminate at approximately 6.1 molecules/nm$^2$, beyond which negative surface energies are observed, violating the assumptions of the Wulff construction. Such a morphology change in MgO nanoparticles has been experimentally observed \cite{geysermans2009combination}. After two days of being immersed in neutral liquid water, initially, cubic MgO nanoparticles exhibited truncations at the cube edges, with \{110\} and \{111\} surfaces becoming visible from different viewing angles. After seven days, many particles adopted a diamond-like shape, consistent with the morphology predicted by our simulations. This indicates that our findings obtained from calculations on surfaces in the vacuum may still be applicable to explain the behaviors of MgO nanoparticles in an aqueous environment. 

With the obtained surface energies, the surface stresses at different surface water coverages were calculated using Eq. (\ref{eq:interfstre}) and shown in Fig. \ref{surf_frac}(d), with negative values meaning lattice expansion at the particle core and positive values meaning lattice contraction at the particle core. For the \{100\} surface, the surface stress stays positive across all water coverage levels, while it stays negative for the \{110\} and \{111\} surfaces. Both \{100\} and \{110\} surfaces exhibit an overall descending trend with the increase of water coverage on the surface, being the opposite to the \{111\} surface. Surface stresses are sensitive to the spatial distribution of water molecules, indicated by the larger error bars than the calculated surface energies. These findings demonstrate the complex interplay between water adsorption and surface stress for different MgO surfaces. The positive surface stresses for the \{100\} surface indicate that, if MgO nanoparticles maintain a cubic shape regardless of water coverage on their surfaces, achieving negative surface stresses would be impossible even with fully hydroxylated surfaces. Thus, we prove that water adsorption and surface hydroxylation alone are not sufficient to induce compressive-to-tensile strain transition, which has been commonly hypothesized in prior works \cite{diehm2012size,chan2021surface}. Instead, surface reconstruction in nanoparticles must occur to reveal facets with negative surface stresses, so that the average surface stress will become negative. 

The relation between surface stress and the $\mathrm{H_2O}$ coverage of the surface can be established by fitting the data in Fig. \ref{surf_frac}(c) with quadratic regressions, and the average surface stress ($f_{ave}$) of a nanoparticle can then be estimated using the rule of mixture:
\begin{equation} 
f_{ave}=\sum_{i=1}^{n}f_{i}\cdot \phi_{i},
\label{eq:ave_stre}
\end{equation} 
where $f_{i}$ and $\phi_{i}$ are the surface stress and surface fraction of individual surfaces at a given water coverage, and $n=3$ is the number of considered facets. Before the water coverage reaches 3.3 molecules/nm$^2$, the average surface stress of a nanoparticle (black curve in Fig. 1(d)) coincides with the surface stress of the \{100\} surface, since nanoparticles are fully bound by these surfaces. With increasing water coverage, the average surface stress decreases rapidly due to the emergence of \{110\} and \{111\} surfaces, which exhibit negative surface stresses. Eventually, the average surface stress matches that of the \{111\} surface, once the nanoparticle surface becomes fully dominated by this facet. An estimation of MgO particle surface stress based on the lattice strain and particle size measured in \cite{rodenbough2017lattice} is approximately -4.13 N/m. Although the water concentration on the surface and the specific surface types were not characterized in the study, TEM images in \cite{rodenbough2017lattice} reveal many particles deviating from a cubic shape, supporting our hypothesis of hydroxylation-induced surface reconstruction and negative surface stress.

\begin{figure}[htbp!]
  \centering
  \includegraphics[width=0.75\textwidth,clip,trim=0cm 0.0cm 0cm 0cm]{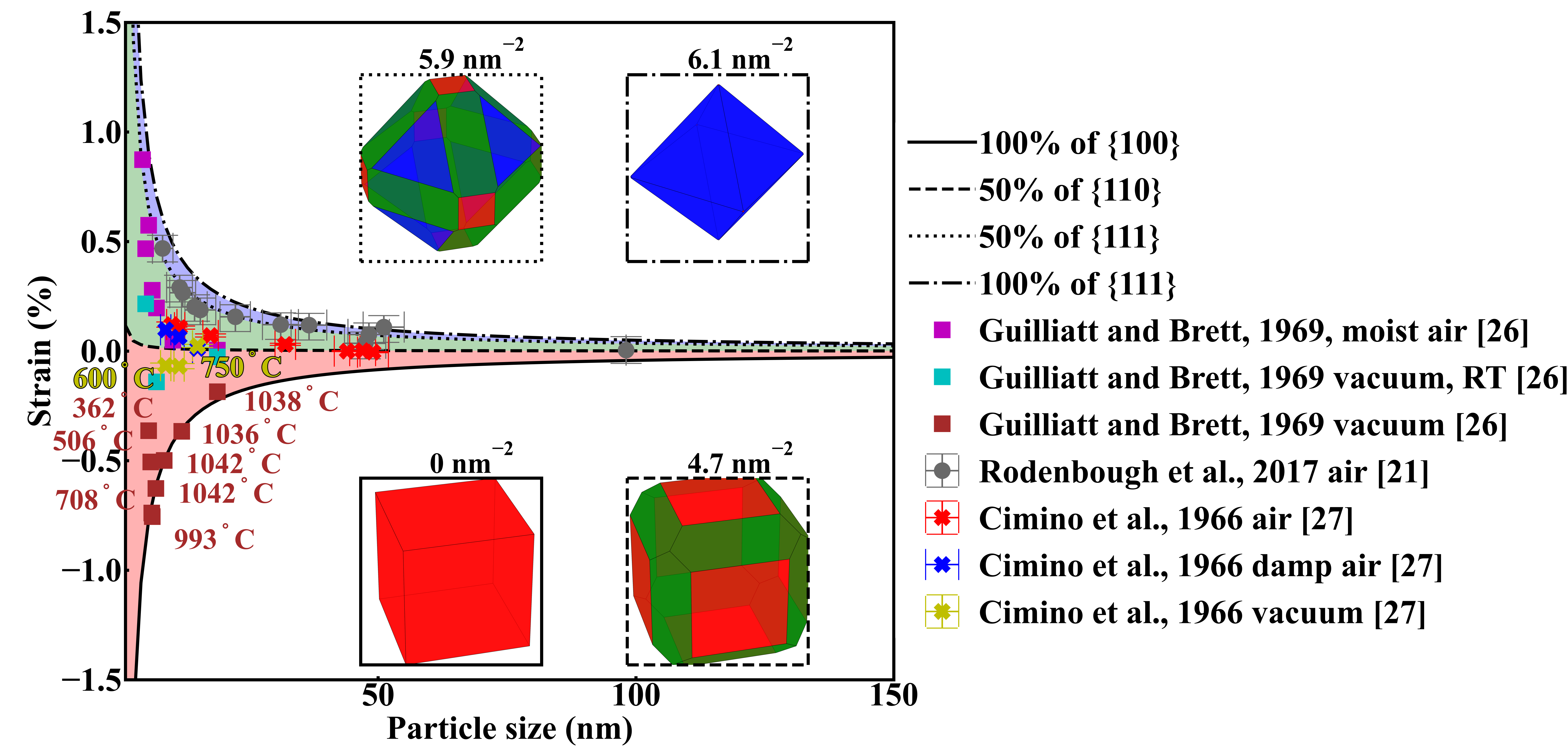}
  \caption{
  The variation of strains within MgO nano-particles with nano-particle sizes. The solid, dashed, dotted, and dash-dotted lines are obtained using the average surface stresses computed for particles with 100 \% of \{100\}, 50 \% of \{110\}, 50 \% of \{111\}, 100 \% of \{111\} surfaces, respectively. The insets show the shapes and surfaces of each particle constructed by WulffPack, with surface water coverage marked on top of each inset. Experimental data are reported in the work of Cimino et al. \cite{cimino1966dependence}, Guilliatt and Brett \cite{guilliatt1969lattice}, and Rodenbough et al. \cite{rodenbough2017lattice}.}
  \label{particle}
\end{figure}

With the predicted surface stresses for nanoparticles at varying water coverages, combined with Eq. (\ref{eq:stre_size}) (which relates lattice strain to particle size and surface stress), we can predict the maximum and minimum lattice strains achievable for particles of specific sizes. As illustrated in Fig. \ref{particle}, a cubic nanoparticle fully bound by \{100\} surfaces and without any water adsorption (solid line) exhibits the most positive surface stress, resulting in the lowest lattice strain. In contrast, a diamond-shaped nanoparticle fully bound by \{111\} surfaces (dash-dotted line) exhibits the most negative surface stress, corresponding to the highest lattice strain. The dashed and dotted lines represent lattice strains for particles with 50\% \{110\} facets and 50\% \{111\} facets, respectively. The regions between these lines are color-coded to indicate dominance by specific surfaces: light red for \{100\}, light green for \{110\}, and light blue for \{111\}.

The experimentally measured lattice strains and particle sizes from \cite{cimino1966dependence,guilliatt1969lattice,rodenbough2017lattice} are plotted with simulation predictions in Fig. \ref{particle}. All experimental data points fall within the upper and lower bounds predicted by our calculations. Notably, several brown squares align perfectly with the solid line representing clean nanoparticles fully enclosed by pristine \{100\} facets. These data correspond to MgO nanoparticles synthesized from $\mathrm{Mg(OH)_2}$ in vacuum and annealed at temperatures ranging from 362 $^\circ$C to 1042 $^\circ$C \cite{guilliatt1969lattice}. Pure MgO nanoparticles were obtained at annealing temperatures at 993 $^\circ$C and above. Below this temperature, water molecules or hydroxyl ions are likely to be retained even under vacuum conditions \cite{anderson1965interaction}, leading to higher-than-predicted lattice constants. For nanoparticles annealed at temperatures over 1000 $^\circ$C but subsequently cooled to room temperature in vacuum, minor moisture reabsorption and subsequent surface reconstruction during the cooling process can slightly raise the lattice constant (cyan squares) \cite{guilliatt1969lattice}. In contrast, nanoparticles prepared under vacuum but later exposed to moist air experience substantial water adsorption, resulting in a significant increase in the lattice constant (magenta squares) \cite{guilliatt1969lattice}.

In \cite{rodenbough2017lattice}, a similar fabrication method is used as in \cite{guilliatt1969lattice}, but MgO nanoparticles were only annealed in air at various temperatures. After cooling down to room temperature, lattice constants and particle sizes were measured using XRD under ambient conditions (grey circles). Water retention and adsorption have likely occurred during both annealing and cooling processes, significantly increasing the measured lattice constant and reaching the extreme positive lattice strains predicted by our model. The predicted surface reconstruction has also been confirmed in \cite{rodenbough2017lattice} by transmission electron microscopy (TEM) for nanoparticles with an average size of 41 nm, exhibiting a non-cubic shape.
In \cite{cimino1966dependence}, MgO nanoparticles were prepared by decomposing magnesium carbonates in air (red crosses) or under vacuum (yellow crosses) at various temperatures, and those maintained under vacuum were later exposed to moist air (blue crosses). The exposure to air in both cases resulted in significant lattice expansion attributable to moisture adsorption as predicted in our work. The possibility of carbon-based product contamination of MgO nanoparticle surfaces in \cite{cimino1966dependence} renders quantitative lattice strain comparison impossible with our work as well as with other experimental works discussed above. Further experimental details from all three works are provided in Section S5.

To summarize, we have shown that water adsorption systematically alters the equilibrium shape and lattice strain of MgO nanoparticles by modifying the surface energy and surface stress. Hydroxylation drives a transition from a positive to negative average surface stress and promotes a shift from the \{100\} facet to the more stable \{110\} or \{111\} surfaces, matching experimental reports of lattice expansion and morphology changes under humid conditions. These simulation results, corroborated by available experimental measurements, illustrate that nanoparticle preparation routes and environmental exposures play a crucial role in determining final particle morphologies.  Our study highlights the importance of combining computational and experimental methods to capture the atomistic details of surface transformations and ultimately harness them for technology-driven applications. Though focused on MgO, our method can be applied to explain the effect of different surface ligands on lattice strain in various nanoparticles. The identified link between facet reconstruction and surface stress suggests a powerful strategy for engineering oxide nanomaterials to meet specific functional requirements. By controlling the water coverage of surfaces and annealing protocols, one can tune active sites for catalysis, modulate interfacial chemistry in sensing and battery applications, and influence nanoparticle interactions in biomedicine. The broad impact of our work is briefly discussed in End Matter. 

\section*{Acknowledgments}
This research was supported by NCCR MARVEL, a National Centre of Competence in Research, funded by the Swiss National Science Foundation (grant number 205602). We also thank Prof. Patrik Hoffmann from Empa/EPFL for the fruitful discussions.

\section*{End Matter}
\paragraph{Applicability to other systems} 
Our findings demonstrate how water adsorption can dramatically alter surface energy, morphology, surface stress, and lattice strain in MgO nanoparticles, ultimately influencing their functional properties. 
Similar lattice expansion in nanoparticles has been observed in a variety of metal oxides such as CeO$_{2}$ \cite{chen2010size,prieur2020size}, rutile TiO$_{2}$ \cite{li2004evidence}, ZnO \cite{li2007p,ali2010zno}, LiCoO$_{2}$ \cite{okubo2007nanosize}, CaWO$_{4}$ \cite{su2007tunable}, MnCr$_{2}$O$_{4}$ \cite{bhowmik2006lattice}, cubic SrTiO$_{3}$ \cite{wu2007negative}, and Ni$_{0.6}$Zn$_{0.4}$Fe$_{2}$O$_{4}$ \cite{naughton2007lattice}, where our findings are highly relevant. The choice of precursors in particle synthesis, which may contain $\mathrm{H_2O}$ (e.g. Mg(NO$_{3}$)$_{2}$·6H$_{2}$O, Zn(NO$_{3}$)$_{2}$·6H$_{2}$O, Na$_{2}$WO$_{4}$·2H$_{2}$O) or OH groups (e.g. Ce(OH)$_{4}$, $\mathrm{Mg(OH)_2}$, Zn(OH)$_{2}$, CoOOH, LiOH), can release $\mathrm{H_2O}$ or OH groups that attach to nanoparticle surfaces during precursor decomposition. Moreover, many syntheses involve heating in open air, allowing moisture from the atmosphere to be adsorbed. The analysis conditions themselves, such as performing XRD under vacuum or under ambient conditions, can also affect residual water uptake. Beyond H$\textsubscript{2}$O, other adsorbates—including CO$\textsubscript{2}$, N$\textsubscript{2}$, CH$\textsubscript{4}$, or functional groups in aqueous solutions—may bind to the nanoparticle surface, introducing distinct surface stresses and corresponding core lattice strains. Although these additional species are not the focus of our current work, measuring lattice strain can nonetheless provide valuable insights into the unique surface chemistry they produce.

Additional factors could contribute to the ultimate lattice expansion of nanoparticles, such as those observed in CeO$_{2}$, where lattice expansion has been linked to surface and subsurface O vacancies and changes in Ce valence \cite{zhou2001size}. Diehm et al. \cite{diehm2012size} supported this idea by reporting a negative surface stress for the non-stoichiometric (111) surface of CeO$_{2}$.
Yet more recently, Prieur et al. \cite{prieur2020size} excluded the link of such a lattice expansion to oxygen vacancies and cation valence using high-resolution X-ray photoelectron spectroscopy (XPS) and high-energy resolution fluorescence-detection hard X-ray absorption near-edge structure (HERFD-XANES) spectroscopy, while highlighting the role of surface hydroxyl and carbonate groups. Our calculations on stoichiometric surfaces demonstrate that for metal oxides such as MgO, surface defects play a negligible role and moisture adsorption alone can adequately explain the observed lattice expansion in experiments.

\paragraph{Characterization breakthrough} 

Historically, XRD measurements of lattice strain in nanoparticles have been conducted sporadically over the past half-century, driven more by researchers’ curiosity and fundamental interest rather than by direct industrial or technological applications. The key challenge for the latter lies in interpreting XRD-measured average lattice strains and reliably correlating them with the nanoparticles' shape, structure, and chemistry. Although the theory of surface stress has existed for decades, experimentally measuring surface stress for a given orientation and surface state remains exceedingly complex, so only the measurements of subsurface residual stresses have been done so far in the literature \cite{cancellieri2021strain,prevey1992problems}. Furthermore, deriving surface stresses from \textit{ab initio} calculations is computationally more demanding than computing surface energies. As shown in this work, the emergence of foundational neural network potentials—which combine the high fidelity of \textit{ab initio} calculations with the computational efficiency required for large-scale simulations enables precise and efficient reproduction of surface properties. Integrating accurate atomistic simulations with XRD and chemical analysis techniques (such as, for example, nuclear magnetic resonance (NMR) and X-ray photoelectron spectroscopy (XPS)) would allow researchers to directly evaluate how even minor variations in water content or other adsorbates influence surface reconstructions in physiological and catalytic environments. 

The synergy between simulations and experiments, as emphasized in \cite{du2022identification}, underscores the broader challenge of characterizing and engineering oxide nanomaterials under non-ideal or ambient conditions. Techniques relying on vacuum or probe molecules provide only indirect glimpses of realistic surfaces, whereas computational models allow for an atomistic perspective on environmental impacts. By aligning theoretical predictions with carefully designed experiments, researchers can systematically validate how water coverage modifies lattice strain and identify which facets are most relevant to real-world applications. This interdisciplinary approach—where computational insights guide experimental design—will likely accelerate the development of advanced oxide nanomaterials for catalysis, sensors, energy storage, and biomedical applications. 

\paragraph{Real-world applications} The potential implications for adsorption and catalysis are especially evident when linking our findings to those of prior studies on oxide nanomaterials. For example, Du et al. \cite{du2022identification} reported that identifying specific oxygen sites on MgO surfaces was critical for understanding CO$_{2}$ adsorption behavior, and the binding of surface oxygen ions with hydrogen ions from the atmosphere can hugely reduce the adsorption of CO$_{2}$ molecules. Our work underscores how water adsorption can tune surface energetics to stabilize or destabilize particular facets, which would exhibit different adsorption sites and adsorption capabilities of water molecules and target molecules, inevitably affecting the adsorption and catalytic efficiency of nanoparticles. Our findings provide valuable insights into nanoparticle synthesis (e.g. the choice of precursors, vacuum conditions, heat treatment, etc.) for advancing heterogeneous catalysis strategies. 

Beyond catalysis, the ability to harness water-driven surface reconstructions has ramifications in sensor \cite{manjunatha2021synthesis,shanawad2023humidity,asha2024improved} and battery \cite{wang2019enhancing,pandey2009magnesium,kumar2007effect} applications as well. MgO-based sensors could become more selective or sensitive if surface stress modifications enhance adsorbate binding, while controlling water adsorption might also mitigate unwanted morphological changes that degrade sensor performance. MgO nanoparticles are also introduced into battery electrodes or electrolytes to enhance their capacitance, conductivity, etc., the interfacial stress can impact both electrochemical stability and ionic transport pathways; designing nanoparticles to present low-stress facets in an operational environment may improve battery cycle life. The knowledge that humid conditions can tip the balance toward certain facets offers a route to engineer stable oxide nanomaterials, with direct connections to manufacturing processes that involve thermal treatments in controlled atmospheres.

Our results also carry significant implications for the field of biomedicine, where nanoparticle surface structure and chemistry dictate interactions with biological molecules and membranes. Metal oxide nanoparticles have already proven promise in antibacterial and anticancer therapies \cite{stoimenov2002metal,huang2005controllable,applerot2009enhanced,cai2017highly,behzadi2019albumin}, in which nanoparticles are usually exposed to an aqueous environment. The nanoparticle morphology change predicted by our simulations, in which the dominant facets of MgO nanoparticles vary with water coverage, suggesting that their ability to interact with bacteria, spores, or cell membranes may also evolve during use. Such morphology change also offers a promising avenue for engineering MgO-based nanostructured materials with enhanced or tunable biological activity, maximizing therapeutic efficacy and biosafety in real-world biomedical applications.

\bibliography{main}

\ifarXiv
    

\section*{Supplementary material}

# Hydroxylation-driven surface reconstruction at the origin of compressive-to-tensile stress transition in MgO nanoparticles

Yang Hu[1,2] and Vladyslav Turlo[*1,2]

[1]*Laboratory for Advanced Materials Processing, Empa - Swiss Federal Laboratories for Materials Science and Technology, Thun, Switzerland*
[2]*National Centre for Computational Design and Discovery of Novel Materials MARVEL, Empa, Thun, Switzerland*
(Dated: March 5, 2025)

## 1. Computational details

Molecular statics calculations were carried out using the Large-scale Atomic/Molecular Massively Parallel Simulator (LAMMPS) [1]. Structural visualization and post-simulation analysis were performed using the Open Visualization Tool (OVITO) [2]. To describe interatomic interactions, a sophisticated universal graph neural network potential (NNP), specifically PreFerred Potential (PFP) was used [3], which covers a wide range of molecular and crystalline systems, supporting simulations involving combinations of up to 96 elements. Its training is based on an extensive dataset comprising over 32 million structures, all consistently generated from high-quality density functional theory (DFT) calculations [4]. To account for weak intermolecular and interatomic interactions, the van der Waals (vdW)-dispersion energy correction proposed by Grimme et al. [5] was applied.

To validate PFP, fundamental materials properties of MgO and $Mg(OH)_2$ were calculated and compared with available DFT and experimental data, including comparisons for lattice constants, bulk modulus, surface energies ({100}, {110}, and {111} surfaces of MgO), adsorption energies for a single $H_2O$ molecule on these surfaces, and geometric parameters such as OH bond lengths and the HOH bond angle in a gas-phase $H_2O$ monomer, as summarized in Table S1 and Table S2.

Table S1 demonstrates that incorporating van der Waals dispersion correction in computations for MgO yields results closer to experimental values for the lattice constant and bulk modulus of MgO, as well as for the {100} surface energy. While those computed without the van der Waals dispersion correction, such as the {100} and {110} surface energies, as well as the adsorption energy of one $H_2O$ molecule on these two surfaces, are very close to the DFT data which were calculated without considering such corrections (those reported in [6]). Table S2 illustrates that for $Mg(OH)_2$, both calculations with and without van der Waals dispersion correction underestimate the lattice constant $c$ and overestimate most of the reported bulk modulus values. The computation with van der Waals dispersion correction provides a lattice constant $a$ closer to experimental data, whereas the calculation without this correction better reproduces the elastic constants of $Mg(OH)_2$.

## 2. The lattice constant of MgO and $Mg(OH)_2$

MgO adopts the Rock-salt structure with eight atoms in a primitive cell (space group: $Fm\bar{3}m$). To obtain the lattice constant for the fully relaxed bulk, a 2×2×1 supercell was created and uniformly strained along all axes to establish the relation between the lattice constant and pressure, see Figure S1. The lattice constant of MgO was then determined by applying a linear regression on the data, which are $\sim 0.42132$ nm and $\sim 0.42578$ nm for calculations with and without van der Waals dispersion corrections, respectively. The obtained lattice constants were used for the following construction of simulation cells for calculating surface energies and stresses. The bulk modulus can also be determined using the variation of pressure with lattice constant, according to $K = V_0 \frac{dP}{dV}$, where $V_0$ is the volume of unstrained cell and $V_0 = 4a^3$.

For $Mg(OH)_2$ ($P\bar{3}m1$), a 6×6×6 supercell was created, with one unit cell containing five atoms. Similarly to MgO, the supercell was uniformly strained along all axes. The variations of pressure with the lattice constants are shown in Figure S2 for both calculations with and without the van der Waals dispersion correction, from which the lattice constant $a$ of a relaxed $Mg(OH)_2$ was determined to be 0.3155 nm and 0.3187 nm, and lattice constant $c$ was determined to be 0.4640 nm and 0.4664 nm. The bulk modulus was computed using $K_V = \frac{C_{11}+C_{22}+C_{33}+2C_{12}+2C_{13}+2C_{23}}{9}$, as reported in [41], with the elastic constants calculated as the coefficients to relate stresses and strains. The results for calculations with and without the van der Waals dispersion correction are determined as 62.160 GPa and 55.547 GPa, respectively (See Table S2 for details).

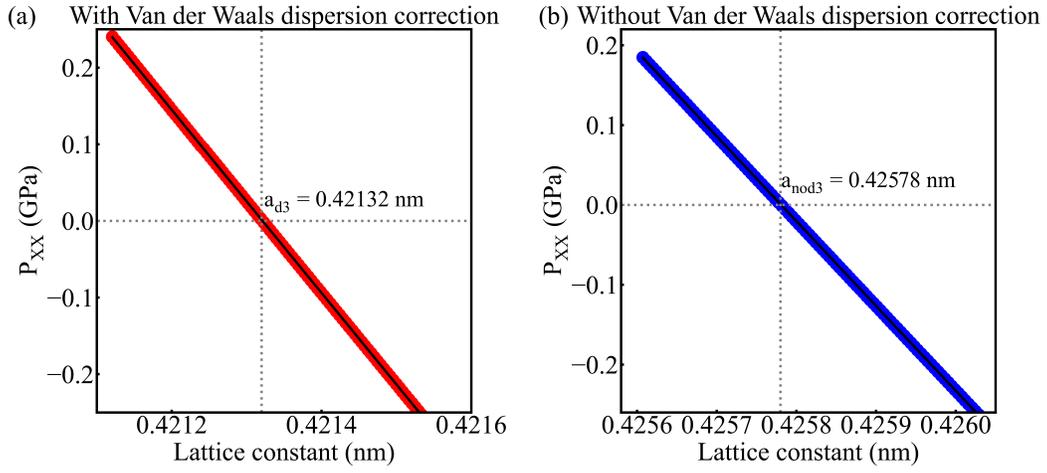

FIG. S1. The variation of pressure along the X-axis with the lattice constant of MgO obtained by calculations (a) with the van der Waals dispersion correction and (b) without van der Waals dispersion correction.

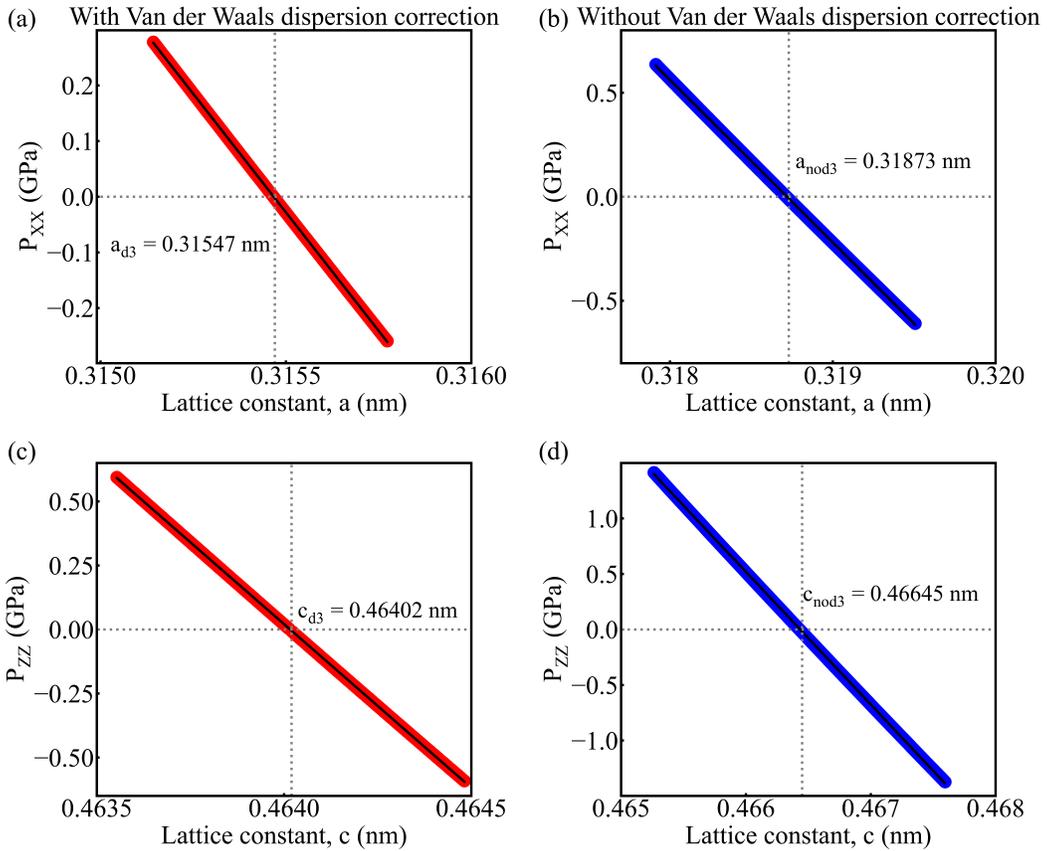

FIG. S2. The variation of pressure along the X-axis with the lattice constant, a, of Mg(OH)$_2$ obtained by calculations (a) with the van der Waals dispersion correction and (b) without van der Waals dispersion correction. The variation of pressure along the Z-axis with the lattice constant, c, of Mg(OH)$_2$ obtained by calculations (c) with the van der Waals dispersion correction and (d) without van der Waals dispersion correction.

### 3. Surface energies and stresses

To calculate the surface energies of {100}, {110}, and {111} surfaces, the simulation cells were oriented with the Z-axis perpendicular to each surface and the separation between the top and bottom surfaces is chosen to be 2-20 nm to test the thickness convergence, as shown in Figure S4. A 0.5-2 nm vacuum layer was added to both the top and bottom surfaces to test the influence of vacuum layer thickness, with periodic boundary conditions being kept along all dimensions. The simulation cell dimensions along the X- and Y-axis for {100}, {110} and {111} surfaces are $0.8426 \times 0.8426$ nm$^2$ ($2a \times 2a$, where a is the lattice parameter of MgO), $0.8426 \times 0.8937$ nm$^2$ ($2a \times 3\sqrt{2}/2a$) and $1.0319 \times 1.1916$ nm$^2$ ($3\sqrt{6}/2a \times 2\sqrt{2}a$), respectively. These surface areas were maintained in the following simulations, and no box relaxation was allowed. Atomic positions were first relaxed with the conjugate gradient (CG) energy minimization method, and then further relaxed with the Fast Inertial Relaxation Engine (FIRE) method [42, 43]. The surface energy was computed from the total system energy $E_{tot}^{surf}(\varepsilon)$ as

$$\gamma(\varepsilon) = \frac{E_{tot}^{surf}(\varepsilon) - E_{tot}^{bulk}(\varepsilon) - n \times E_{H_2O}}{2A(\varepsilon)} \tag{1}$$

where $E_{tot}^{surf}(\varepsilon)$ and $E_{tot}^{bulk}(\varepsilon)$ are the potential energies of simulation cells with and without surfaces at biaxial in-plane strain, $\varepsilon$, and $2A$ is the total surface area. $n$ is the number of $H_2O$ molecules and $E_{H_2O}$ is the energy of a $H_2O$ monomer in gas phase. For the case of the {111} surface, we chose the one on the top to be O-terminated, and the one at the bottom to be Mg-terminated, so that the net charge in the simulation is zero. The computed surface energy is then the average over these two types of surfaces.

The surface stress $f(\varepsilon)$ under a biaxial in-plane strain, $\varepsilon$, is related to the surface energy, $\gamma(\varepsilon)$, by the Shuttleworth equation

$$f(\varepsilon) = \gamma(\varepsilon) + \frac{\partial \gamma(\varepsilon)}{\partial \varepsilon} \tag{2}$$

The value of $f(\varepsilon)$ can thus be determined by computing surface energy versus biaxial strain, as recently performed by density-functional theory (DFT) calculations for idealized Cu/W interface structures [44].

One example of the calculated surface energies and stresses for the {100}, {110} and {111} surfaces of MgO is shown in Figure S3.

### 3.1. The effect of vacuum layer thickness on surface energies

The computed surface energies of {100}, {110}, and {111} surfaces with a 10 nm surface separation and different vacuum layer thicknesses are shown in Figure S4. The surface energies increase with the vacuum layer thickness and reach asymptotic values after the vacuum layer thickness reaches 1 nm, meaning that 1 nm thickness is enough to avoid the interaction between surfaces through the vacuum layer. In the following simulations, a 1.5 nm thickness for the vacuum layer is used.

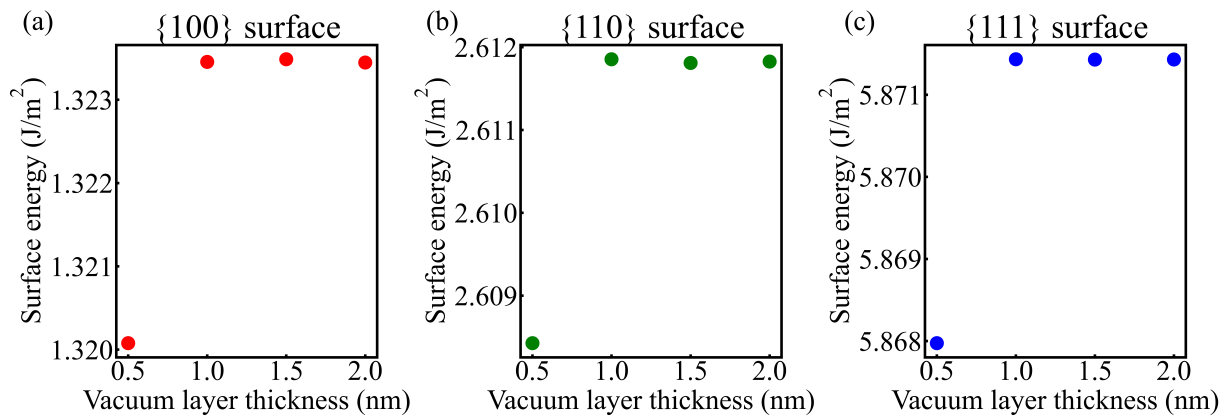

FIG. S4. The variation of (a) {100}, (b) {110} and (c) {111} surface energies with the vacuum layer thickness.

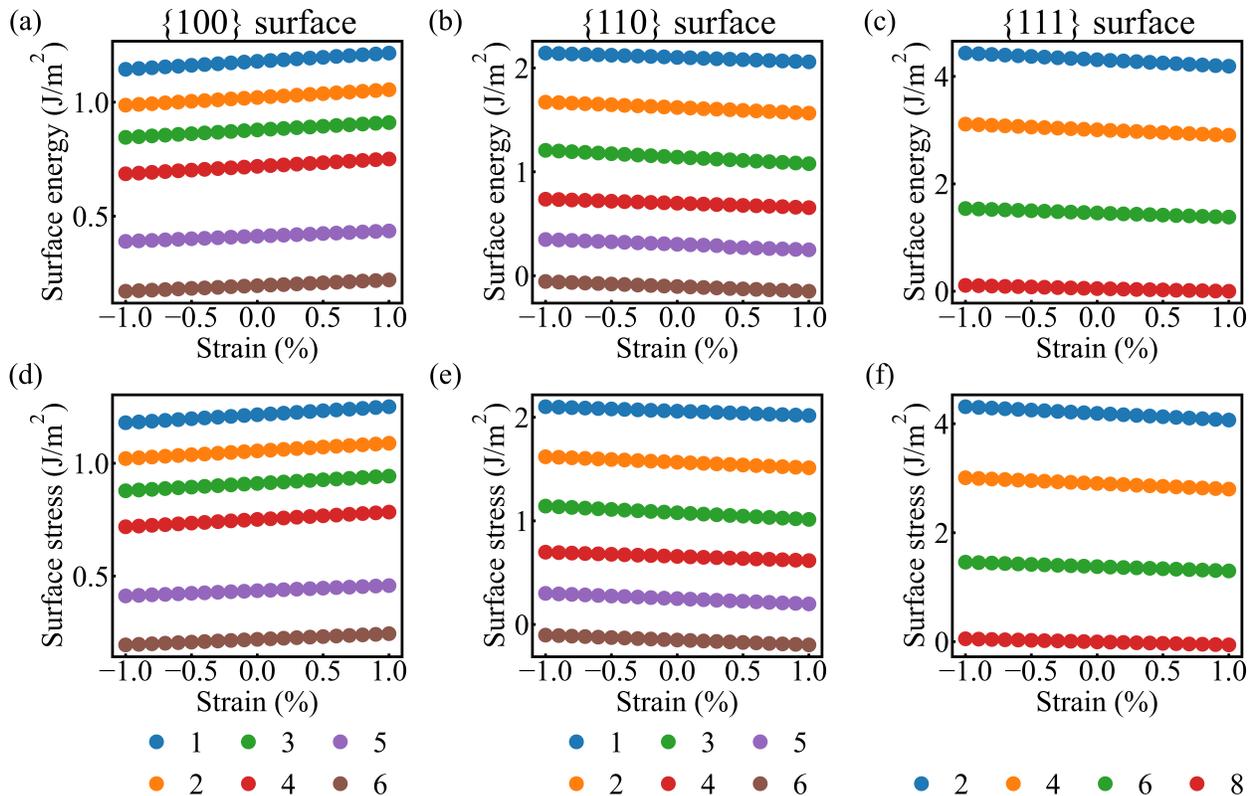

FIG. S3. The variation of surface energies and stresses with biaxial strains for the {100}, {110}, and {111} surfaces of MgO. Different curves represent different numbers of water molecules adsorbed on the surface.

### 3.2. The effect of surface separation on surface energies

The computed surface energies of {100}, {110}, and {111} surfaces with varying surface separations are presented in Figure S5. As observed for all three surfaces, increasing the surface separation leads to the surface energy converging to an asymptotic value. Similar behavior has been documented in various face-centered cubic and body-centered cubic metals, as reported in [45]. There are multiple contributions to the computed excess energy of the slab compared to the bulk system. Firstly, atoms near the surface exhibit a lower coordination number than those in the bulk, contributing to excess energy. Secondly, strain energy in the sub-surface atomic layers induced by surface stresses also affects the computed energy, diminishing as surface separation increases. Lastly, strong interactions between the two surfaces are notable at smaller separations.

In our simulations, the in-plane dimensions were set based on the equilibrated bulk lattice constant, and box relaxation was not allowed, thereby eliminating the influence of the second contribution. With increasing surface separation, the interaction between the top and bottom surfaces diminishes, resulting in surface energies stabilizing at asymptotic values for sufficiently large separations. Figure S5 demonstrates that a surface separation of 10 nm is adequate to minimize inter-surface interactions, and this value will be utilized for subsequent simulations.

The calculated surface stresses at a 10-nm surface separation for the 100, 110, and 111 surfaces are 5.226 J/m$^2$, -0.420 J/m$^2$, and -8.302 J/m$^2$, respectively. One report on the {100} surface stress calculated using DFT is 0.223 eV/Å$^2$=3.57 J/m$^2$ [46], not too far from our computed value. The deviation could originate from the structure optimization method and the simulation size.

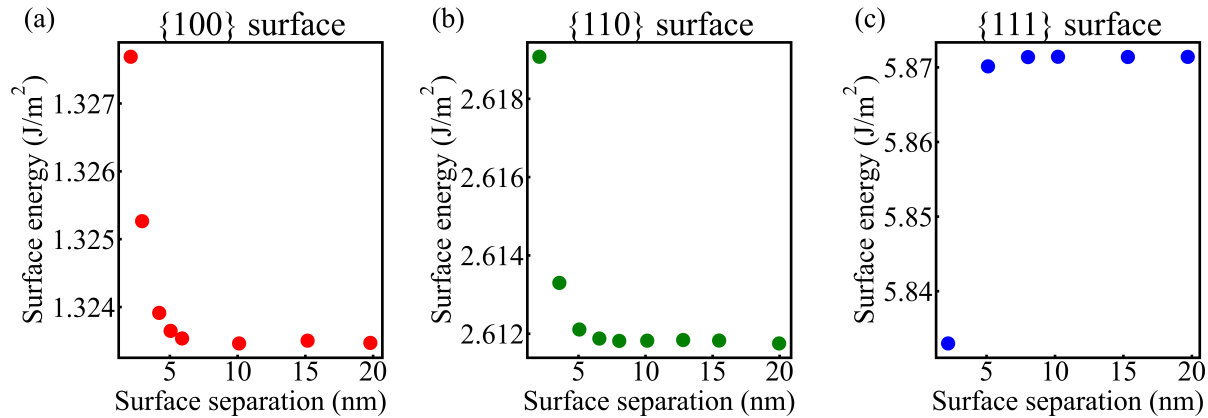

FIG. S5. The variation of (a) {100}, (b) {110} and (c) {111} surface energies with the surface separation.

### 3.3. The effect of surface water coverage on surface energies and stresses

To investigate the effect of $H_2O$ coverage on the surface energies and stresses of various MgO surfaces, different numbers of $H_2O$ molecules were introduced to the top and bottom surfaces of the slabs. For the {100} and {110} surfaces, one OH group was positioned 1 Å above or below a randomly selected Mg atom on the top or bottom surface, while one H atom was placed 1 Å above a neighboring oxygen atom. For the {111} surface, one OH group was positioned 1 Å below a randomly selected Mg atom from the Mg-terminated surface, and one H atom was placed 1 Å above a randomly selected oxygen atom from the oxygen-terminated surface. Atomic positions were again optimized using the CG energy minimization method [47], followed by the FIRE method [42, 43]. Box dimensions were fixed to maintain a consistent surface area across simulations, ensuring a fair comparison regardless of the number of $H_2O$ molecules.

For a 2a × 2a MgO {100} surface, there are eight Mg and eight O atoms available for bonding with OH groups and H atoms, as illustrated in Figure S6. For a 2a × 3√2/2a MgO {110} surface, six Mg and six O atoms are available (Figure S7), while a 3√6/2a × 2√2a MgO {111} surface features sixteen Mg or sixteen O atoms per plane (Figure S8). For a given $H_2O$ coverage, multiple configurations of $H_2O$ molecule distributions are possible. The configurations used for surface energy and stress calculations in this study are shown in Figure S6-Figure S8. While not all possible configurations were evaluated for the {111} surface due to the numerous variations, the selected configurations are considered representative and provide reliable average values for both surface energies and stresses.

### 4. The adsorption energy of $H_2O$ on MgO surfaces

To obtain the adsorption energy of one $H_2O$ molecule on various MgO surfaces, the simulation cells for calculating the surface energies are expanded four times for {100} and {110} surfaces, and three times for the {111} surface to avoid the interaction between the inserted water molecule and its periodic image. To introduce one $H_2O$ molecule on the {100} and {110} surfaces, one OH group was positioned 1 Å above a randomly selected Mg atom on the top surface, then one hydrogen atom was positioned 1 Å above a neighboring oxygen atom to the Mg atom. For the {111} surface, one OH group was positioned 1 Å below a randomly selected Mg atom from the Mg-terminated surface and one hydrogen atom was positioned 1 Å above a randomly selected oxygen atom from the oxygen-terminated surface. The structures were then relaxed with the CG and FIRE methods, and the simulation box relaxation along the X-axis and Y-axis was allowed. The relaxed configurations are shown in Figure S11, with the bond lengths of OH groups marked in each frame. A $H_2O$ monomer under vacuum was also simulated with the OH bond length comparable to reported experimental and DFT data, see Table S1. The $H_2O$ monomer on the {100} surface is not dissociated, and the bond length for the OH group that is not interacting with the surface is maintained as 0.97 Å, while the one interacting with the surface becomes ~1.0083 Å, a value similar to what is reported in [48]. However, in the scenario of high surface water coverage and $H_2O$ molecules located close to each other (see Figs. S9 and S10), the dissociation of water molecules on the {100} surfaces can occur and the surface energy can be lower than those with no dissociation of water molecules.

By contrast, the $H_2O$ monomer on the {110} surface is dissociated, and the OH group interacting with the surface is stretched and becomes ~1.5914 Å. Such a phenomenon has also been observed in [6].

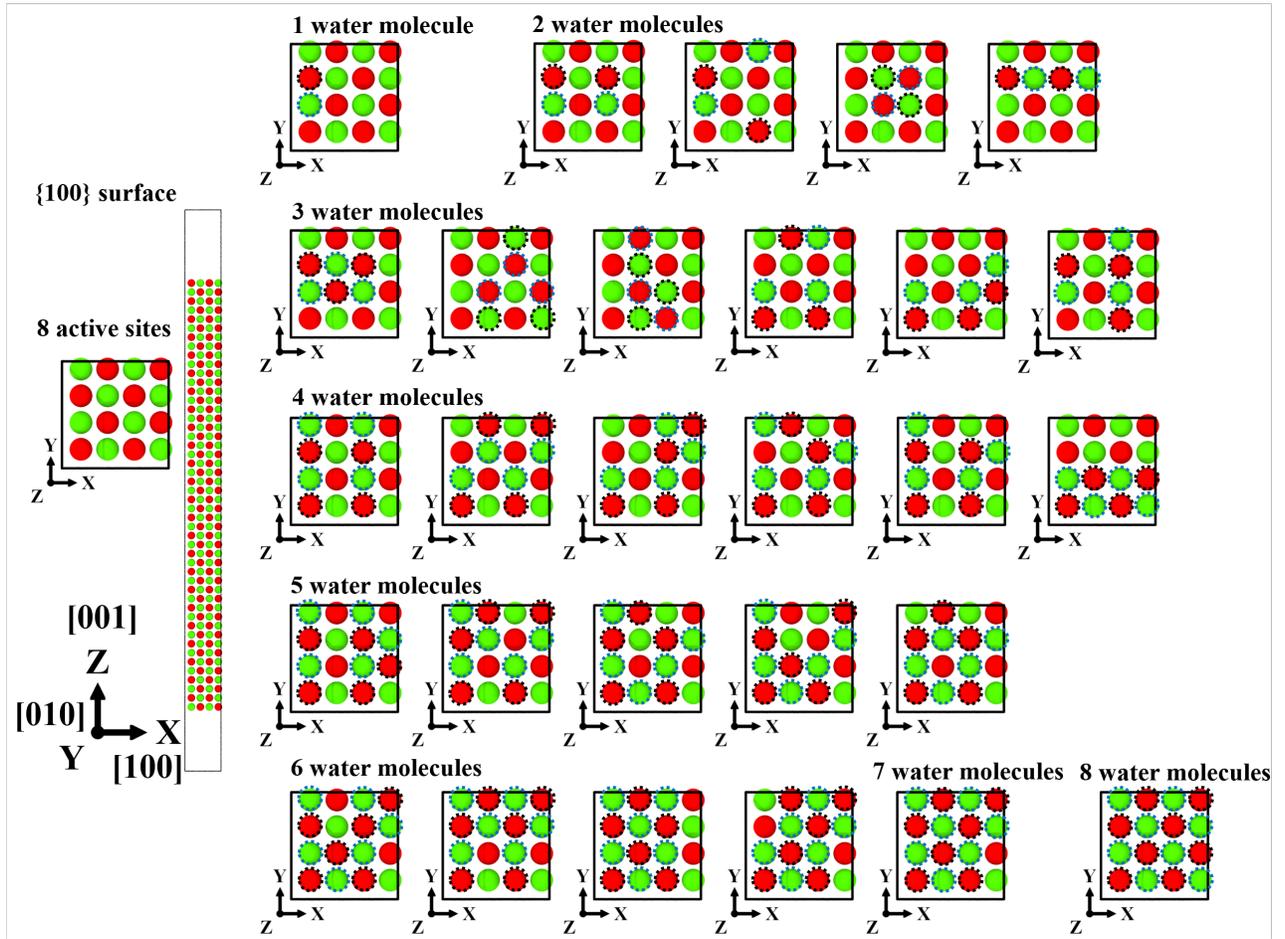

FIG. S6. The possible positions of water molecules on the {100} surface. The positions of Mg atoms above which the OH groups were positioned are highlighted using blue dashed circles, and the positions of O atoms above which the H atoms were positioned are highlighted using black dashed circles.

The adsorption energy, $E_{ads}$, is calculated as

$$E_{ads} = E_{hydro,surf} - E_{surf} - n \times E_{H_2O} \tag{3}$$

where $E_{hydro,surf}$ and $E_{surf}$ are the energies of the simulation box with and without $H_2O$ molecules on the surface, respectively. $n$ is the number of $H_2O$ molecules and $E_{H_2O}$ is the energy of a $H_2O$ monomer in gas phase.

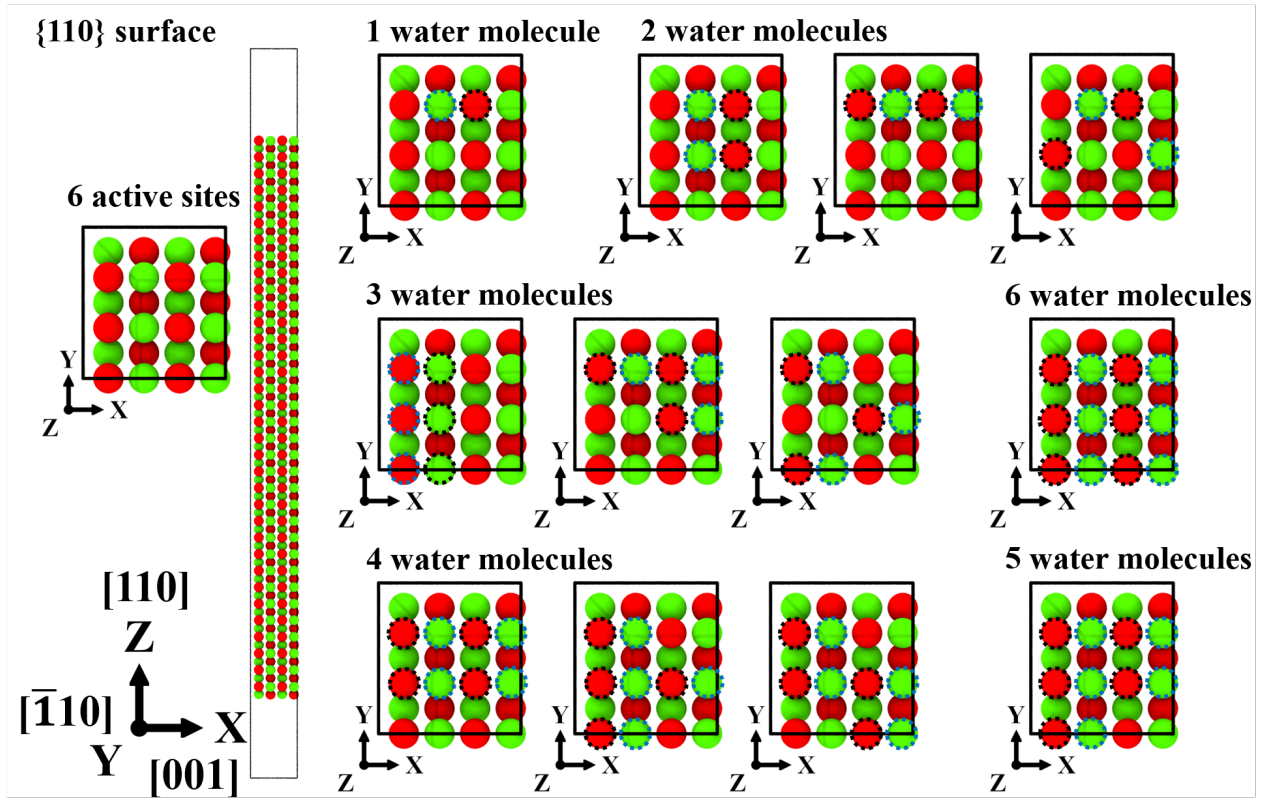

FIG. S7. The possible positions of water molecules on the {110} surface. The positions of Mg atoms above which the OH groups were positioned are highlighted using blue dashed circles, and the positions of O atoms above which the H atoms were positioned are highlighted using black dashed circles.

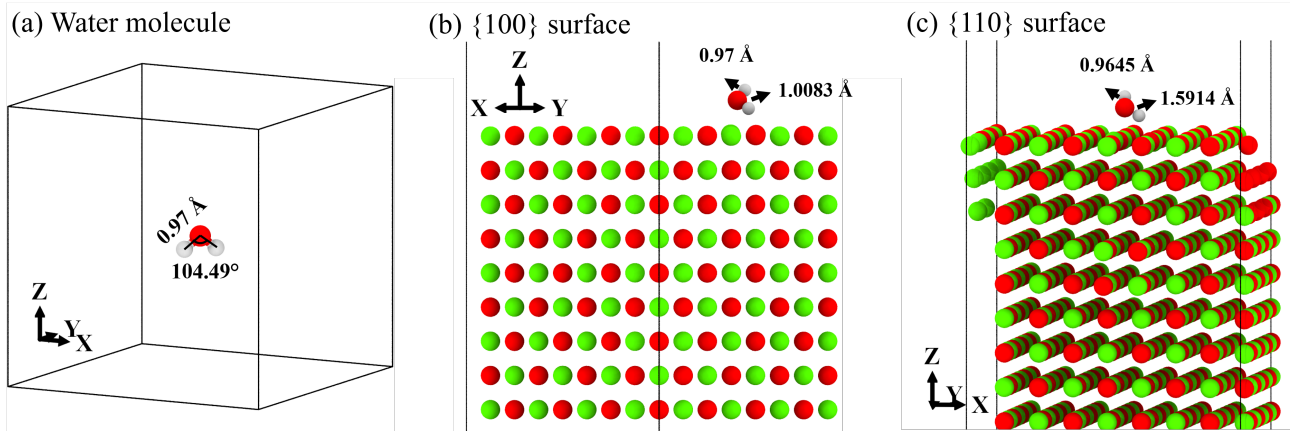

FIG. S11. The relaxed configuration of (a) a $H_2O$ monomer, (b) a $H_2O$ monomer on the {100} surface, and (c) a $H_2O$ monomer on the {110} surface. Mg atoms are colored green, oxygen atoms are colored red, and hydrogen atoms are colored grey. The bond lengths of the OH groups are marked in (a)-(c).

## 5. Experimental details

In [10], MgO nanoparticles were prepared through three different methods: (1) pure MgO obtained from Johnson, Matthey & Company, Limited (London) (denoted as MgO-JM); (2) MgO obtained by decomposition of magnesium carbonate in air at 500–1000 °C (MgO-C, purple crosses); and (3) MgO from magnesium carbonate decomposition in

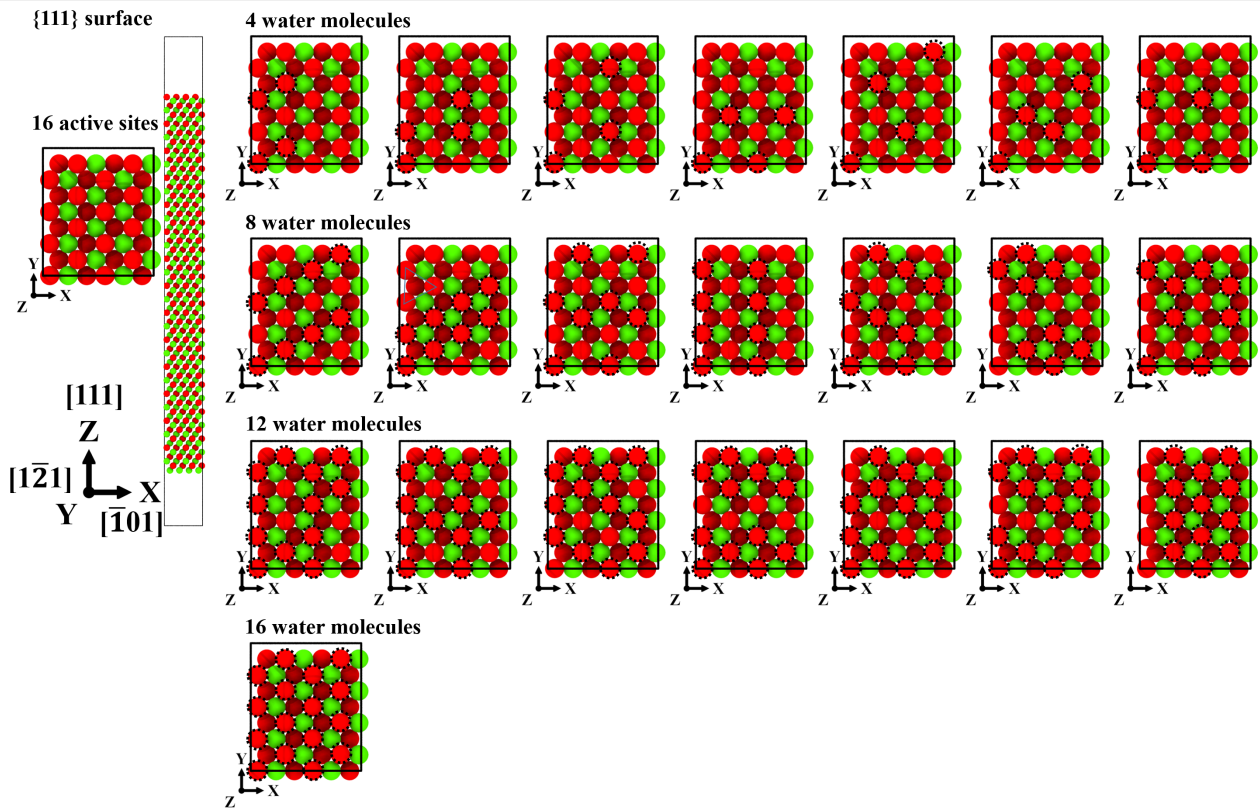

FIG. S8. The possible positions of water molecules on the {111} surface. The positions of O atoms above which the H atoms were positioned are highlighted using black dashed circles. The distribution of OH groups below the Mg-terminated surface is the same.

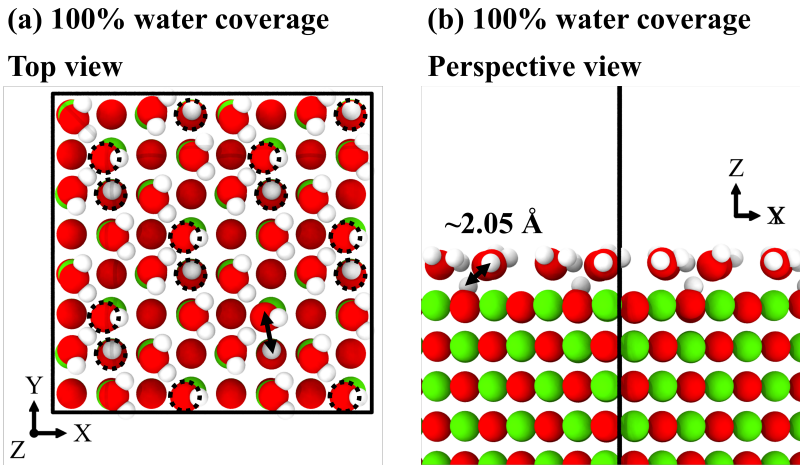

FIG. S9. The (a) top view and (b) perspective view of water molecules on {100} surfaces of MgO for 100% water coverage. The dashed circles in (a) mark the water molecules that dissociate, and the distance between the O atom in the water and the H atom interacting with the surface is ~2.05Å, being much bigger than the OH bond length in an undissociated water.

vacuum at 600–900 °C (MgO-CV). (1) was fired in air at temperatures from 450 to 1300 °C (red crosses) or in H$_2$ at 800 and 1000 °C (orange crosses). (3) was later exposed to damp air at room temperature (blue crosses). Lattice constants were measured at various temperatures and adjusted to 21 °C using a thermal expansion coefficient of $1.3 \times 10^{-5}$/°C. The bulk lattice constant of MgO was determined to be 4.2115 Å from high-temperature measurements of MgO-JM and MgO-C.

**(a) Distribution 1**        **(b) Distribution 2**

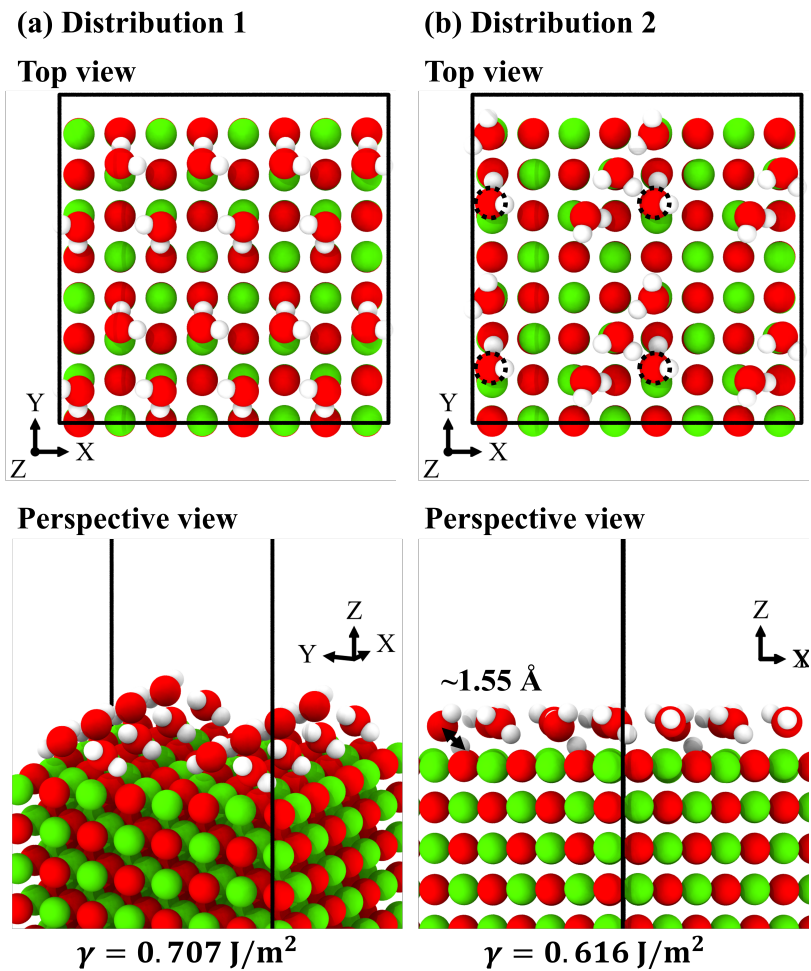

FIG. S10. The top view and perspective view of water molecules on {100} surfaces of MgO for 50% water coverage with two different spatial distributions of water molecules. The dashed circles in (b) mark the water molecules that dissociate, and the distance between the O atom in the water and the H atom interacting with the surface is ~1.55Å, being much bigger than the OH bond length in an undissociated water, yet close to the OH bond lengths of dissociated water molecules on the {110} surface.

In [9], $Mg(OH)_2$ powder was decomposed at 325 °C overnight and further heated in vacuum to temperatures from 362 to 1038 °C (brown squares). Some powders were cooled to room temperature (canyon squares) and later exposed to moist air (magenta squares). Lattice parameters were corrected to 23 °C using thermal expansion coefficients for MgO of $13.63 \times 10^{-6}$/°C (25–1000 °C) and $12.83 \times 10^{-6}$/°C (25–500 °C). The bulk lattice constant, determined from hydroxide calcined at 1550 °C for 70 hours, was reported as $4.211 \pm 0.003$ Å at 23 °C.

In [49], MgO nanoparticles were synthesized from magnesium nitrate hexahydrate and annealed in air at 500–1100 °C for 2 hours (grey circles). Lattice constants were measured via XRD under ambient conditions, and the bulk lattice constant was reported as 4.2050 Å.

The measured lattice constants and particle sizes from [9, 10, 49] are shown in Tables S3-S6 below.

Further comparison reveals that the vacuum-prepared particles in [10] tend to have larger average lattice constants than those reported by Guilliatt and Brett [9], who employed more critical experimental conditions. Specifically, Cimino et al. [10] used a vacuum of $10^{-5}$ mmHg at 600-900 °C for 5 hours, whereas Guilliatt and Brett first decomposed $Mg(OH)_2$ at 325 °C overnight under $10^{-6}$ mmHg, followed by heating at various temperatures for over 16 hours. Cimino et al.[10] also examined how vacuum quality ($10^{-1}$ vs. $10^{-5}$ mmHg) and annealing duration (5 h vs. 20 h) influenced the final lattice constant, observing smaller values for higher-vacuum, longer-annealed samples. These findings suggest that the MgO nanoparticles in [9] achieved much higher purity and effectively minimized

surface contamination, consistent with their more rigorous vacuum and annealing protocols.

––––––––––––––––––––––––

TABLE S1. Materials properties of MgO

| Properties | Here | Experiment | Computation |
|---|---|---|---|
| Lattice constant (nm) | 0.42132 0.42578* | 0.4213 [7], 0.42117 [8], 0.42113 [9], 0.42115 [10] | 0.4259 [11], 0.42506 [12], 0.4215 [13], 0.4165 [14], 0.4217±0.0001 [15] |
| Bulk modulus (GPa) | 167.7186 150.7477* | 160 [7], 160.2 [8] | 145.68 [11], 159.7 [12], 160.9 [13], 171 [14], 162±2 [15] |
| {100} surface energy (J/m$^2$) | 1.3232 0.8549* | 1.04 [16], 1.2 [17], 1.15±0.08 [18], 1.28 [19], 1.37 [19] | 1.43 [20], 1.17 [21], 0.85 [6], 1.10 [15] |
| {110} surface energy (J/m$^2$) | 2.6123 2.0699* | | 2.05 [6], 2.29 [21] |
| {111} surface energy (J/m$^2$) | 5.8715 5.1905* | | 5.78 [22] |
| Adsorption energy of H$_2$O on {100} surface (eV/atom) | -0.6329 -0.6092* | | -0.51 [6], -0.537 [23], -0.4976 [24], -0.477 [25] |
| Adsorption energy of H$_2$O on {110} surface (eV/atom) | -2.5221 -2.4215* | | -2.36 [6] |
| Adsorption energy of H$_2$O on {111} surface (eV/atom) | -6.0233 −5.8111* | | |
| H$_2$O bond length (Å) | 0.9700 0.9700* | 0.9572 [26] | 0.966 [15], 0.959 [27], 0.97 [28], 0.97-0.973 [28], 0.958-0.962 [28] |
| HOH bond angle (deg) | 104.490 104.490* | 104.52 [26] | 103.9-104.1 [15], 104.2 [27], 104.9 [28], 104.1-104.4 [28], 104.8-105.1 [28] |
| The formation energy of one isolated H$_2$O molecule in gas phase (eV) | -10.0887 −10.0789* | | |

Note: *data calculated without van der Waals corrections. Fei et al. [7] employed synchrotron X-ray diffraction and hydrostatic compression to determine these properties at room temperature. Similarly, Speziale et al. [8] utilized a comparable approach. Guilliatt [9] calcined $Mg(OH)_2$ at 1550 °C for 70 hours to produce MgO, measuring the lattice constant at high temperatures and correcting it to 23 °C. Cimino et al. [10] fired MgO obtained from Johnson, Matthey & Company, Limited at temperatures exceeding 1000 °C and determined the lattice constant via X-ray analysis, converting results to 21 °C. Refson et al. [15] and Gueddim et al. [11] computed these properties using ab initio methods based on the local-density approximation (LDA) and the generalized gradient approximation (GGA), respectively. Karki et al. [12] and Zhao et al. [13] employed pseudopotential methods within the LDA and norm-conserving pseudopotentials in the Kleinman–Bylander representation with the GGA exchange-correlation function, respectively. Baltache et al. [14] utilized a non-scalar relativistic FP-LAPW approach within density functional theory (DFT), applying the LDA functional by Perdew and Wang.

Surface energies of MgO were estimated by Jura [16] based on heats of solution and heat capacities, and by Gilman et al. [17] using liquid nitrogen cleavage. Gutshall [19] and Westwood [18] employed similar cleavage techniques in different conditions. Causa et al. [20] computed 100 surface energies using an ab initio Hartree-Fock crystalline orbital LCAO program, while Goniakowski et al. [21] utilized a quantum self-consistent method with geometry optimization. Chen et al. [6] investigated surface energies and H$_2$O adsorption energies using the GGA-PBE functional and ultrasoft pseudopotentials. Gaddy et al. [22] employed the PAW method and PBEsol functional to average Mg-terminated and O-terminated surface energies. Ončák et al. [23] used DFT-PBE+D2 for surface energy calculations, while Carrasco et al. [24] and Hu et al. [25] utilized the PAW method with various exchange-correlation potentials. The OH bond length and HOH bond angle were determined by Benedict et al. [26] through rotation-vibration spectra of deuterated H$_2$O vapor. Tschumper et al. [27] computed these parameters using a triple-ζ basis set augmented with polarization and diffuse functions, optimizing geometries at the CCSD(T) theoretical level. Xu et al. [28] investigated bonding using multiple GGA, LDA, and hybrid GGA functionals within DFT.

TABLE S2. Materials properties of $Mg(OH)_2$

| Properties | Here | Experiment | Computation |
|---|---|---|---|
| Lattice constant, a (nm) | 0.3155<br>0.3187* | 0.314±0.0002 [29], 0.3145±0.0001 [30], 0.314979±0.000004 [31], 0.31425±0.0002 [32], 0.31472±0.00002 [33], 0.314±0.001 [34], 0.315±0.0001 [35] | 0.3189 [36], 0.3126 [36], 0.3148 [37], 0.3179 [37], 0.3167 [37], 0.3134±0.0004 [38] |
| Lattice constant, c (nm) | 0.4640<br>0.4665* | 0.4769±0.0002 [29], 0.4769±0.0001 [30], 0.47702±0.000001 [31], 0.47665±0.0003 [32], 0.47699±0.00011 [33], 0.476±0.001 [34], 0.4783±0.0005 [35] | 0.4773 [36], 0.4730 [36], 0.5263 [37], 0.4717 [37], 0.4854 [37], 0.4663±0.0001 [38] |
| $C_{11}$ | 161.767<br>156.370* | 156.7±0.8 [34], 159.0±1.4 [35] | 156.3 [36] |
| $C_{12}$ | 46.384<br>43.241* | 44.4±1 [34], 43.3±1.7 [35] | 45.0 [36] |
| $C_{13}$ | 14.748<br>11.563* | 12.0±1.5 [34], 11.1±2.5 [35] | 10.1 [36] |
| $C_{14}$ | 0.392<br>−0.454* | 0.2±0.8 [34], 1.3±1 [35] | 0.2 [36] |
| $C_{33}$ | 65.564<br>62.089* | 46.3±0.8 [34], 49.5±0.7 [35] | 50.4 [36] |
| $C_{44}$ | 32.609<br>23.334* | 21.7±0.5 [34], 22.8±0.4 [35] | 21.8 [36] |
| Bulk modulus (GPa) | 62.160<br>55.547* | 51±4 [39], 54.3±1.5 [29], 47±5 [40], 42±2 [30], 41±2 [31], 44±1 [32], 41.8±1.3 [33], 36.7±1 [34], 39.6±1.4 [34], 35.8±0.9 [35] | 46.4 [36], 47.5±0.7 [38] |

*Note: *data calculated without van der Waals corrections. The lattice constants for brucite reported in [29, 30, 32, 33, 35] were obtained using powder X-ray diffraction patterns. Catti et al. [31] employed neutron diffraction and Rietveld refinement to obtain the lattice constant at room pressure. Xia et al. [34] utilized X-ray diffraction on a single crystal of natural brucite. Pishtshev et al. [36] computed lattice constants using GW-versions of PAW-PBE pseudopotentials and PBE GGA exchange-correlation functions, as well as with the HSE06 hybrid functional. Ugliengo et al. [37] reported lattice constants using HF, PW91, and B3LYP hybrid exchange functionals.*

*The bulk modulus of brucite has been determined using various techniques. Duffy [39] employed shock wave compression, while Fei et al. [29, 32, 33] used static compression and powder X-ray diffraction. Parise et al. [31, 40] also used static compression and neutron diffraction. Duffy [30] and Jiang [35] utilized single crystal X-ray diffraction and Brillouin scattering, respectively, to determine bulk modulus and elastic constants. Xia et al. [34] obtained bulk modulus values through Brillouin scattering and fitting diffraction-measured lattice parameters with the Birch-Murnaghan equation of state. In this work, the bulk moduli were obtained using $K_V = \frac{C_{11}+C_{22}+C_{33}+2C_{12}+2C_{13}+2C_{23}}{9}$, as reported in [41].*

TABLE S3. Values of, respectively: annealing temperatures, lattice constants, and particle sizes reported in [10]

| T (K) | Lattice constants (Å) | Particle sizes (Å) |
|---|---|---|
| MgO-JM | | |
| 450 | 4.2138±0.0005 | 164±10 |
| 600 | 4.2137±0.0002 | 197±15 |
| 700 | 4.2133±0.0001 | 250±15 |
| 800 | 4.2124±0.0001 | 475±25 |
| 900 | 4.2118±0.0001 | 625±25 |
| 1000 | 4.2116±0.0001 | 820±40 |
| 1200 | 4.2115±0.0001 | 1250±80 |
| 1300 | 4.2115±0.0001 | 1470±100 |
| MgO-C | | |
| 500 | 4.2166±0.0005 | 100±10 |
| 550 | 4.2160±0.0003 | 115±10 |
| 600 | 4.2145±0.0003 | 175±15 |
| 700 | 4.2128±0.0002 | 320±20 |
| 800 | 4.2116±0.0001 | 470±25 |
| 900 | 4.2115±0.0001 | 440±25 |
| 1000 | 4.2113±0.0001 | 490±30 |
| MgO-JM fired in $H_2$ | | |
| 800 | 4.2118±0.0001 | 570±30 |
| 1000 | 4.2114±0.0001 | 1050±80 |

TABLE S4. Values of, respectively: vacuum conditions, annealing temperatures, annealing time, lattice constants, and particle sizes reported in [10]

| Vacuum conditions (torr) | T (K) | t (h) | Lattice constants (Å) | Particle sizes (Å) |
|---|---|---|---|---|
| MgO-CV | | | | |
| $10^{-5}$ | 600 | 5 | $4.2087 \pm 0.0005$ | $88 \pm 10$ |
| After moist air | | | 4.2156 | $88 \pm 10$ |
| $10^{-5}$ | 750 | 5 | $4.2086 \pm 0.0005$ | $114 \pm 10$ |
| After moist air | | | 4.2142 | $114 \pm 10$ |
| $10^{-5}$ | 900 | 5 | $4.2125 \pm 0.0003$ | $150 \pm 10$ |
| After moist air | | | 4.2120 | $150 \pm 10$ |

TABLE S5. Lattice constants and particle sizes measured in vacuum at various temperatures [9].

| T (K) | Lattice constants (Å) | Particle sizes (Å) |
|---|---|---|
| 362 | 4.1960 | 55 |
| 506 | 4.1900 | 59 |
| 708 | 4.1802 | 61 |
| 993 | 4.1795 | 62 |
| 1042 | 4.1849 | 69 |
| 1042 | 4.1903 | 85 |
| 1036 | 4.1959 | 119 |
| 1038 | 4.2035 | 188 |

TABLE S6. Lattice constants and particle sizes measured at room temperature [9].

| Lattice constants (Å) | Particle sizes (Å) |
|---|---|
| Moist air | |
| 4.2481 | 43 |
| 4.2310 | 49 |
| 4.2355 | 55 |
| 4.2230 | 62 |
| 4.2196 | 70 |
| 4.2134 | 102 |
| 4.2115 | 188 |
| No moist air | |
| 4.2204 | 49 |
| 4.2054 | 70 |
| 4.2102 | 188 |

TABLE S7. Values of, respectively: annealing temperatures, lattice constants, and particle sizes reported in [49]. Error bars of lattice constants were estimated by , and those of particle sizes were estimated by .

| T (K) | Lattice constants (Å) | Particle sizes, XRD (nm) |
|---|---|---|
| 500 | 4.2247 | 8.2 |
| 550 | 4.2172 | 11.5 |
| 600 | 4.2162 | 12.1 |
| 650 | 4.2135 | 14.4 |
| 700 | 4.2129 | 15.5 |
| 750 | 4.2116 | 22.3 |
| 800 | 4.2101 | 31.1 |
| 850 | 4.2100 | 36.6 |
| 900 | 4.2096 | 51.1 |
| 950 | 4.2082 | 48.2 |
| 1000 | 4.2065 | 41.5 |
| 1100 | 4.2052 | 98.1 |
| 1200 | 4.2050 | >100 |


\fi

\end{document}